\documentclass[twocolumn,superscriptaddress,showpacs,preprintnumbers,amsmath,amssymb,prl,floatfix]{revtex4}

\usepackage{graphicx}
\usepackage{dcolumn}
\usepackage{latexsym}

\newcommand{\amp}[2]{\ensuremath \left\langle #1\right|T\left|#2\right\rangle}
\newcommand{\An}{\mbox{${\rm A}{_{N}}$}}
\newcommand{\ASS}{\ensuremath{{\rm A}{_{SS}}}}
\newcommand{\ANN}{\ensuremath{{\rm A}{_{NN}}}}
\newcommand{\ALL}{\ensuremath{{\rm A}{_{LL}}}}
\newcommand{\ASL}{\ensuremath{{\rm A}{_{SL}}}}

\newcommand{\Aall}{\ANN, \ASS, and \ASL}
\newcommand{\grad}{\ensuremath{{^{\circ}}}}
\newcommand{\chiz}{\ensuremath{\chi^2}}
\newcommand{\ocite}[1]{Ref.~\protect\onlinecite{#1}}

\newcommand{\ISKP}{Helmholtz-Institut f\"ur Strahlen- und Kernphysik, Universit\"at Bonn,
D-53115 Bonn, Germany}
\newcommand{\HH}{Institut f\"ur Experimentalphysik, Universit\"at Hamburg,
D-22761 Hamburg, Germany}
\newcommand{\IKP}{Institut f\"ur Kernphysik, Forschungszentrum J\"ulich,
D-52425 J\"ulich, Germany}
\newcommand{\ket}[1]{\ensuremath{\left| #1 \right\rangle }}

\begin{document}

\preprint{nucl-ex/0209006}

\title{Measurement of Spin Correlation Parameters \ANN, \ASS, and \ASL\
at 2.1~GeV in Proton-Proton Elastic Scattering}
\author{ F.~Bauer}\affiliation{\HH} 
\author{ J.~Bisplinghoff}\affiliation{\ISKP} 
\author{ K.~B\"u\ss{}er}\affiliation{\HH} 
\author{ M.~Busch}\affiliation{\ISKP} 
\author{ T.~Colberg}\affiliation{\HH} 
\author{ L.~Demir\"ors}\affiliation{\HH}
\author{ C.~Dahl}\affiliation{\ISKP}
\author{ P.D.~Eversheim}\affiliation{\ISKP} 
\author{ O.~Eyser}\affiliation{\HH}
\author{ O.~Felden}\affiliation{\IKP}  
\author{ R.~Gebel}\affiliation{\IKP} 
\author{ J.~Greiff}\affiliation{\HH} 
\author{ F.~Hinterberger}\affiliation{\ISKP}
\author{ E.~Jonas}\affiliation{\HH} 
\author{ H.~Krause}\affiliation{\HH}
\author{ C.~Lehmann}\affiliation{\HH}
\author{ J.~Lindlein}\affiliation{\HH}
\author{ R.~Maier}\affiliation{\IKP} 
\author{ A.~Meinerzhagen}\affiliation{\ISKP}
\author{ C.~Pauly}\affiliation{\HH}
\author{ D.~Prasuhn}\affiliation{\IKP}
\author{ H.~Rohdje\ss}\email{rohdjess@iskp.uni-bonn.de}\affiliation{\ISKP}
\author{ D.~Rosendaal}\affiliation{\ISKP}
\author{ P.~von~Rossen}\affiliation{\IKP}
\author{ N.~Schirm}\affiliation{\HH}
\author{ W.~Scobel}\affiliation{\HH} 
\author{ K.~Ulbrich}\affiliation{\ISKP}
\author{ E.~Weise}\affiliation{\ISKP}
\author{ T.~Wolf}\affiliation{\HH}
\author{ R.~Ziegler}\affiliation{\ISKP}
\collaboration{EDDA Collaboration}

\date{\today}

\begin{abstract}
At the Cooler Synchrotron COSY/J\"ulich spin correlation
parameters in elastic proton-proton (pp) scattering 
have been measured with a 2.11~GeV
polarized proton beam and a polarized hydrogen atomic beam target. 
We report results for \ANN, \ASS, and \ASL\ for c.m. scattering angles
between 30\grad\ and 90\grad. Our data on \ASS\ -- the first
measurement of this observable above 800~MeV -- clearly disagrees with
predictions of available of pp scattering phase shift
solutions while \ANN\ and \ASL\ are reproduced reasonably well.
We show that in the direct reconstruction of the scattering amplitudes
from the body of available pp elastic scattering data at 2.1 GeV the
number of possible solutions is considerably reduced.

\end{abstract}
\pacs{24.70.+s, 25.40Cm, 13.75Cs, 11.80Et, 29.20.Dh, 29.25.Pj   }
\keywords{Polarization, Spin Correlations, Nucleon-Nucleon}

\maketitle
The nucleon-nucleon (NN) interaction as one of the fundamental
processes in nuclear physics has been studied over a broad energy
range and its contribution to our understanding of the strong interaction
cannot be overestimated. NN elastic scattering data, parameterized by
energy-dependent phase-shifts, are used as an important ingredient
in theoretical calculations of inelastic processes, nucleon-nucleus and
heavy-ion reactions. Below the pion production threshold at about
300~MeV elastic scattering is described to a high level of
precision \cite{Machleidt:2001rw} by a number of models,
e.g. phenomenological and meson 
exchange. Here, also effective field theory
\cite{Bedaque:2002mn} has recieved increased attention. An
unambiguous determination of phase shift parameters has been achieved
up to about 0.8~GeV
\cite{Stoks:1993tb,Bystricky:1987,Bystricky:1990,Nagata:1996tp,Arndt:2000xc}.
However, with increasing energy the number of partial waves to be
determined grows, but the quality and density of the experimental data
base diminishes. Recently, it has been pointed out
\cite{Bystricky:1998rh,Arndt:2000xc} that above about 1.2~GeV serious
discrepancies between Phase-Shift Analysis (PSA) of different groups
exist. A model independent Direct Reconstruction of the Scattering
Amplitudes (DRSA) by the Saclay-Geneva group \cite{Bystricky:1998rh}
has not solved this puzzle, since usually both PSA solutions are
supported by one of the two or three solutions found.

Apparently, this issue can only be settled by new, precise
experimental data, especially in observables not measured to date in
this energy domain. Storage rings offer  an unique
environment to perform internal high-statistics experiments with pure
polarized hydrogen targets and polarized beams
\cite{meyer97}. Measurements of spin correlation parameters have been
pioneered  by the PINTEX group at IUCF
\cite{Haeberli:1997,Rathmann:1998,vonPrzewoski:1998ye} at energies
between 200 and 450~MeV. 
In the same spirit the EDDA-collaboration has made a first measurement
of the spin-correlation parameters \ANN, \ASS, and \ASL\ at 2.11~GeV (In Argonne-notation
\cite{Bourrely:1980mr}) at COSY/J\"ulich \cite{Maier:1997}. 
For \ASS\ this is the first measurement above 0.8~GeV and challenges
existing PSA and DRSA predictions, which vary considerably.

\begin{figure}
\caption{Schematic overview of the atomic beam target (top) and the
EDDA-detector (bottom)}
\label{fig:exp}
\includegraphics[width=8cm]{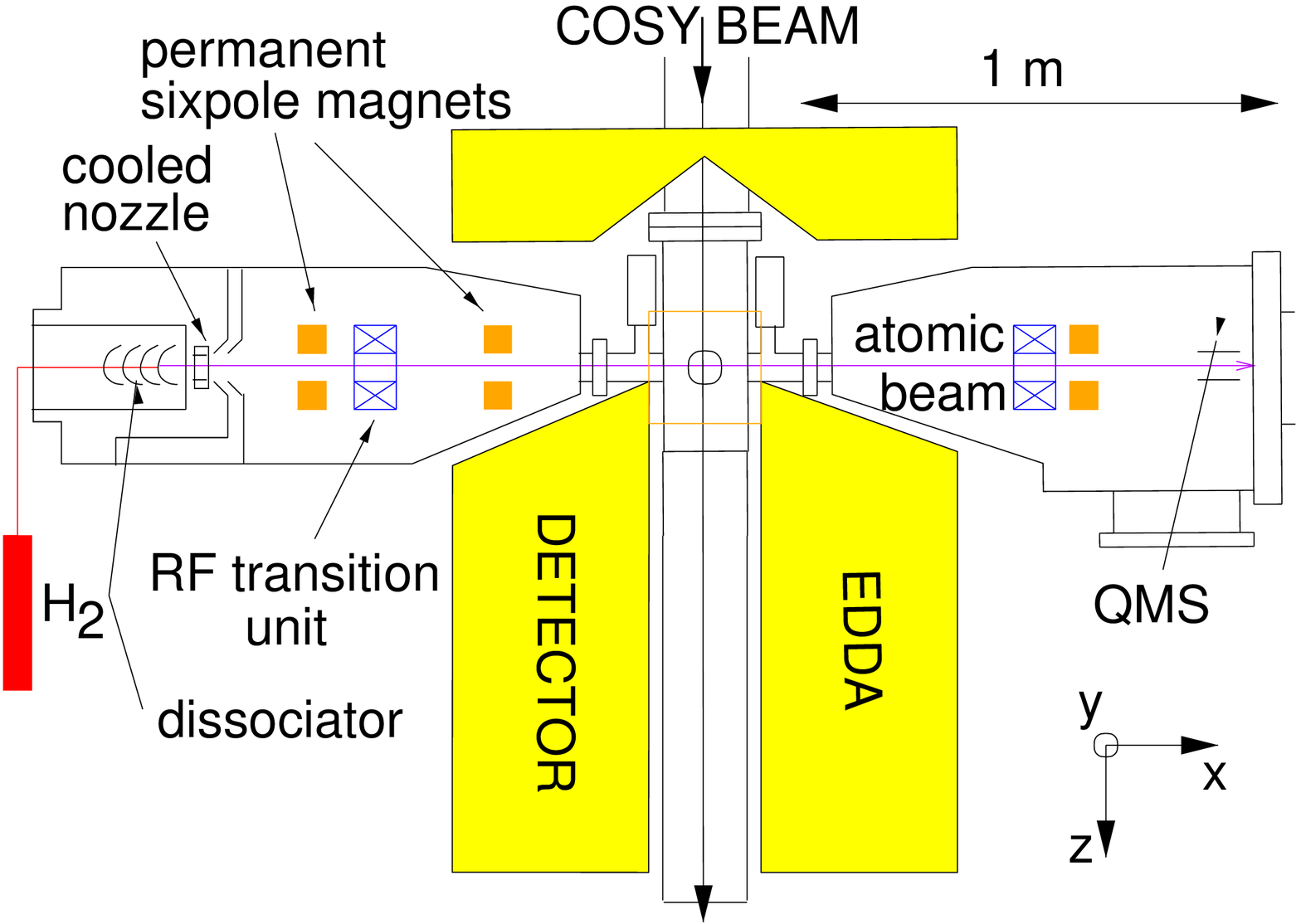}\\
\includegraphics[width=8cm]{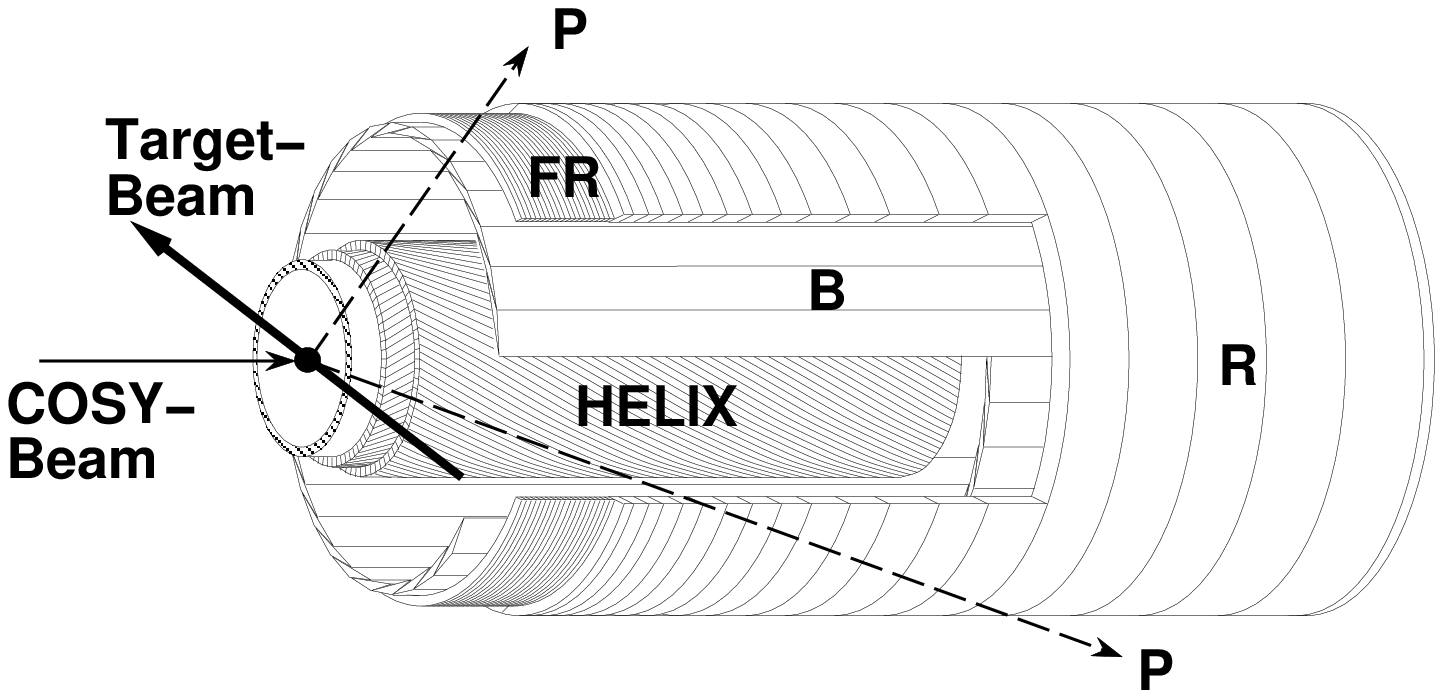}
\end{figure}

The EDDA experiment is set-up at an internal target station at the
COSY ring. The polarized target \cite{eversheim97}, a standard atomic beam
source, is shown schematically in Fig.~\ref{fig:exp}. Molecular hydrogen
is dissociated in a RF-discharge and a atomic hydrogen beam is formed
by a cooled nozzle. By two permanent sixpole magnets and an
RF transition unit only one hyperfine state -- where both proton and 
electron have magnetic quantum number $+\frac{1}{2}$ -- is focused 
into the interaction region, where it crosses the COSY proton
beam. After being analyzed in a Breit-Rabi-spectrometer with respect
to the polarization the hydrogen beam is removed in a beam dump.
In the interaction region the polarization of the hydrogen atoms is
aligned by applying a weak (1~mT) magnetic guiding field in either
one of 6 possible directions $\pm x$, $\pm y$, and $\pm z$. 
Data are acquired after acceleration of the COSY beam at 2.11~GeV for
6~s. After each COSY machine cycle, where the target polarization is
held constant,  the polarization of the COSY beam
($\pm y$) is flipped. Scattering data for the resulting 12 spin
combinations are detected by a double-layered cylindrical scintillator
hodoscope, the EDDA detector (cf. Fig.~\ref{fig:exp} and Refs.~\onlinecite{albers97,Altmeier:2000}).

Event reconstruction proceeds in three steps: First the position of
hits in the inner and outer detector layer are determined. Most events
show multiplicities compatible with two charged particles. Scattering
vertex coordinates and scattering angles are then obtained by 
fitting two tracks to the observed hit pattern. For accepted elastic
events the angular resolution is increased by adding 
constraints imposed by elastic scattering kinematics in the fit.

\begin{figure}
\caption{Distribution of the kinematic deficit $\alpha$ for two ranges
of c.m. scattering angles. Events with $\alpha > 6.25$\grad , mainly due
to inelastic processes or secondary reactions of the final-state
protons in the detector, are
discarded.} 
\label{fig:alpha}
\includegraphics[width=7cm]{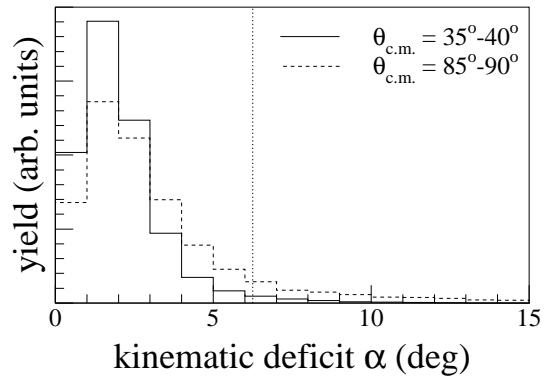}
\end{figure}

Elastic scattering events were selected as described in
\cite{Altmeier:2000}. Two classes of cuts are applied: first, to
match the reaction volume to the beam-target overlap, and second, to
reduce unwanted background contributions from inelastic reactions.
Cuts on the scattering vertex along the COSY-beam  (-15~mm $< z <$
20~mm) are chosen to fully include the width of the atomic beam target
(13~mm FWHM). Perpendicular to the COSY-beam axis, e.g. in the $x-y$
plane, the COSY beam location and width (standard deviation
$\sigma$) can be reconstructed. Events with a reaction
vertex outside a 3$\sigma$-margin of the beam position are discarded.
Elastic scattering events are characterized that the two protons are
emitted back to back in the c.m. system. A cut on the deviation from this
perfect 180\grad\ angle, the so-called kinematic deficit
(Fig.~\ref{fig:alpha}), removes 
inelastic reactions very effectively. However, Monte-Carlo studies
have shown that contributions of three-body final states at the level
of a few percent cannot be eliminated. Their relative contribution can
be varied by loosening or tightening the selection criteria within
reasonable limits, we found no change in the resulting spin
correlation parameters within the statistical errors. 
The arising absolute systematic uncertainty is typically 0.01.

The spin dependence of the NN interaction leads to a modulation of the
cross section $\sigma$ with the azimuthal angle $\phi$, which depends
on the alignment of the beam ($\vec{P}$) and target ($\vec{Q}$)
polarizations with respect to the scattering plane
\cite{bystricky78,Bourrely:1980mr,vonPrzewoski:1998ye}. For the beam
polarization parallel to $y$ one obtains: 

\begin{equation}\label{eq:modul}
\begin{array}{rlll}
\frac{\sigma\left(\theta,\phi,\vec{P},\vec{Q}\right)}{\sigma_0}\  = &
1 &  + \\
 \An(\theta) &  \left\{(P_y+Q_y)\cos\phi - Q_x
\sin\phi\right\} & + \\ 
 \ASL(\theta) &\left\{P_y Q_z \sin\phi \right\} & + \\ 
 \ANN(\theta)& \left\{P_yQ_y\cos^2\phi -
 P_y Q_x\sin\phi\cos\phi\right\}  & + \\
  \ASS(\theta)&  \left\{P_yQ_y\sin^2\phi + P_y
Q_x\sin\phi\cos\phi\right\}.
\end{array}
\end{equation}
Here, $\sigma_0$ is the unpolarized differential cross section,
$\theta$ the c.m. scattering angle and $\An$ the analyzing power,
which is known \cite{Altmeier:2000} and used as input to fix the
overall polarization scale. 

Apart from the spin correlation parameters and polarizations the
integrated luminosities $L$ and detection efficiencies $\epsilon$ must 
be extracted from the measured number of scattering events:
\begin{equation}
N(\theta,\phi,\vec{P},\vec{Q}) = \sigma(\theta,\phi,\vec{P},\vec{Q})\ 
L(\vec{P},\vec{Q})\  \epsilon(\theta,\phi).
\end{equation}
We have employed two methods which
yield consistent results. First, from the sum of events in four quadrants 
in $\phi$, asymmetries in the spirit of \cite{ohlsen73} are defined
which are insensitive to $L$ and $\epsilon$ \cite{Bauer:2001}, or,
secondly, these are determined by a fit using standard \chiz\ minimization
techniques for the set of equations (1) for the 12 spin combinations.
The beam and target polarizations obtained from the data are
$61.1\pm1.5$\% and $69.7\pm1.8$\%, respectively. Measurements of the
magnetic field distribution (modulus and direction) in the interaction
region and a 
determination of unwanted magnetic field components from the data by
the $\chi^2$-fit show that misalignment of the target holding field are
very small and have negligible effect on the spin correlation
parameter.

\begin{figure}[thb]
\caption{Results for \Aall\ in comparison to
PSA-predictions from the Virginia \cite{Arndt:2000xc} (solution FA00,
solid) and  Saclay-Geneva \cite{Bystricky:1998rh} 
(dashed lines) groups as well as data from
SATURNE (\ANN : \protect\cite{Lehar:1987pw,Allgower:2000pv,Allgower:2001qs},
\ASL : \protect\cite{Perrot:1988tw,Fontaine:1989ak,Allgower:1998wf})}
\label{fig:res}
\includegraphics[width=8cm]{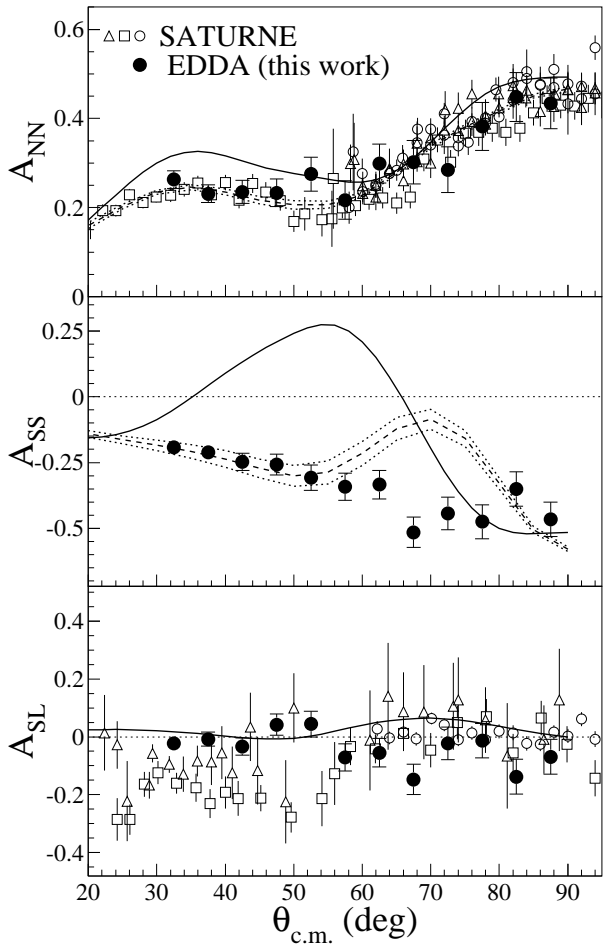}
\end{figure}

The experimental data have been binned in 5\grad\ wide bins in the
c.m. scattering angle $\theta$. The results for \Aall\ are displayed
in Fig.~\ref{fig:res}. An overall normalization uncertainty of 4.7\%, not
included in the error bars,
arises from the statistical and normalization uncertainty of the
analyzing power data \cite{Altmeier:2000} used in the analysis. 
The new data on \ANN\ are in good agreement with
previous measurements and recent PSA solutions. However, for
most of the angular range \ASS\ is in striking disagreement with both PSA
predictions, except for $\theta = 90\grad$ where basic symmetry
considerations require $\ASS\ = 1 - \ANN\ - \ALL$ , with the
right-hand side known
experimentally. 
For \ASL , nearly consistent with zero, we find 
agreement with \cite{Fontaine:1989ak,Allgower:1998wf} but cannot
confirm the results of Ref.~\onlinecite{Perrot:1988tw}, which are
significantly  below zero at small angles. 
Our data are available upon request.

Apparently, available phase-shift parameters
\cite{Bystricky:1998rh,Arndt:2000xc} have failed strongly,
when confronted with data on the observable \ASS\ not present in the
PSA databases. Preliminary analysis of data taken at eight other
energies between 1.35 and 2.5 GeV in subsequent running periods show
the same result, with the dicrepancies gradually increasing
with energy. That phase-shift parameters may not yet be uniquely
determined by experimental data has been pointed out previously
\cite{Bystricky:1998rh,Arndt:2000xc} and would explain naturally the
differences in the PSA-predictions for \ASS\ in Fig~\ref{fig:res}.
To this end a direct reconstruction of the scattering amplitudes
(DRSA), model-independent by design, prooves to be a useful tool.

Exploiting basic symmetry-principles of the strong interaction, the
transition matrix $T$ for pp elastic scattering is given by five
complex amplitudes \cite{bystricky78}. In the
helicity-frame, where \ket{+} and \ket{-} are the positive or negative
helicity states of the protons in the c.m., these five amplitudes are
\cite{JacobWick}
\begin{equation}
\begin{array}{l@{\ \ \ \ \  }l}
\phi_1=\amp{++}{++} & \phi_2=\amp{++}{--}\\
\phi_3=\amp{+-}{+-} & \phi_4=\amp{+-}{-+}\\
\phi_5=\amp{++}{+-},
\end{array}
\label{eq:hel}
\end{equation}
which correspond to no ($\phi_{1,3}$), 
single ($\phi_{5}$) and double-spinflip ($\phi_{2,4}$).
All observables can be expressed by bilinear combinations \cite{Bourrely:1980mr}
of these amplitudes, e.g. for \ASS\  we obtain
\begin{equation}
\begin{array}{rcl}
\ASS  \sigma_0 &=& {\rm Re}(\phi_1\phi_2^*) +{\rm
Re}(\phi_3\phi_4^*) \\
=& \multicolumn{2}{l}{
\left|\phi_1\right|\!\left|\phi_2\right|\cos\left(\alpha_1\!-\!\!\alpha_2\right)\! +\! 
\left|\phi_3\right|\!\left|\phi_4\right|\cos\left(\alpha_3\!-\!\!\alpha_4\right)
}
\end{array}
\label{eq:amp}
\end{equation}
when using $\phi_k =
\left|\phi_k\right|\exp\left(i\alpha_k\right)$.
Given experimental data on at least nine suitably chosen, linearly
independent observables at the same beam energy and scattering angle,
the helicity amplitudes are obtained by a simple \chiz-minimization,
except for an arbitrary, unobservable global phase.
In addition, the choice of helicity-amplitudes sheds light on contributions
of the various spin-dependent parts of the NN-interaction 
\cite{Conzett:1994rg} to the observables. \ASS\ in particular 
depends on the interference between double- and non-spinflip
amplitudes and thus probes aspects of the spin-spin
and spin-tensor parts of the NN interaction.
  
\begin{figure}
\caption{DRSA solutions for some helicity amplitudes without (open) and with (solid) inclusion of the new
data on \ANN, \ASS, and \ASL\  in comparison to PSA solutions 
(as in Fig.~3).
All amplitudes
have been divided by $\sqrt{2\sigma_0} = \left(\sum_{i=1}^5|\phi_i|^2+3|\phi_5|^2\right)^{1/2}$.
The symbols ({\Large $\bullet$},$\blacksquare$,$\blacktriangle$,$\blacktriangledown$, and {\large
$\circ$},$\Box,\triangle,\diamondsuit$ respectively)
distinguish DRSA solutions, corresponding to different minima of the
$\chi^2$-function. 
}
\label{fig:ampl}
\begin{center}
\includegraphics[width=8.5cm,bb=29 6 525 370]{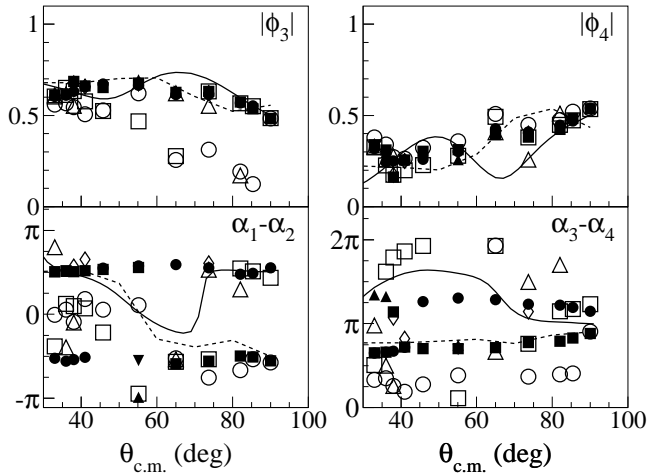}
\end{center}
\end{figure}

We have performed a DRSA at 2.1~GeV at 11 angles, using the
database described in Ref.~\onlinecite{Bystricky:1998rh} with the
addition of recently published data 
on \An\ and ${\rm K}_{0nn0}$
\cite{Ball:1999cn,Ball:1999yy,Altmeier:2000}.
Differential cross section data except those of \ocite{albers97} were
removed. In accordance with \cite{Bystricky:1998rh} we found two to
four minima in the contour of the \chiz-function for a fit of the data
in terms of the amplitudes $\phi_{1\ldots 5}$ (open symbols in
Fig~\ref{fig:ampl}). Thus, the helicity amplitudes are not uniquely
determined, since different sets describe the data equally well.

In order to explore to what extent the inclusion of the new spin
correlation data of this work are a remedy, we have added these to the
database and repeated the search for minima in the \chiz-function
(solid symbols in Fig.~\ref{fig:ampl}).
First, the number of ambiguities is reduced to two in most cases.
Secondly, the \chiz\ is not increased 
by the new data on \ASS , although here, the aforementioned conflict of
Ref.~\onlinecite{Perrot:1988tw} with our data on \ASL\ contributes.
This demonstrates that the new data on \ASS\ are perfectly compatible with
all experimental information available and provide important
additional constraints in determining scattering amplitudes and 
phase-shifts.

For some helicity-amplitudes the different solutions are displayed in
Fig.~\ref{fig:ampl}. The absolute values of all amplitudes are very
well determined when taking the data of this work, in particular on
\ASS , into account. The remaining ambiguities are in the relative
phases of these amplitudes. 
Our DRSA yields that $\left|\alpha_1-\alpha_2\right|\approx\frac{\pi}{2}$, such
that $Re(\phi_1\phi_2^*)$ vanishes. Therefore in Eq.~\ref{eq:amp},
\ASS\ is only driven by $\phi_3$ (no spin flip) and $\phi_4$ (double
spin flip) for antiparallel spins in the initial state. In the PSA
solutions $\cos(\alpha_1-\alpha_2)$ does not vanish and they
overestimate \ASS\ in the  $\theta=50\grad-70\grad$ range. However, some
aspects of the DRSA amplitudes are remarkably well represented by PSA
predictions, like $\alpha_3-\alpha_4$ by the solution of
\cite{Bystricky:1998rh}. 
From the identity
$(\ANN + \ASS)\sigma_0 = 2 Re(\phi_1\phi_2^*) +
2\left|\phi_5\right|^2$ \cite{Bourrely:1980mr}
and the experimental result $\ANN \approx -\ASS$ we conclude further,
that the single spin-flip amplitude 
$\phi_5$, mainly driven by spin-orbit forces \cite{Conzett:1994rg}, must
be small at these energies.

We have reported the first results of elastic $\vec{p}\vec{p}$
scattering at COSY. Spin correlation parameters \Aall\ have been
determined at 2.11~GeV between 30\grad\ and 90\grad\ in the
center-of-mass. While \ANN\ and \ASL\ are at least in reasonable
agreement with previous data and PSA solutions, \ASS\ as the first
measurement above 0.8~GeV adds genuine new
information to the field. 
Therefore, current PSA solutions should be used with caution above 1.2~GeV.
Including these new data has reduced
the ambiguities in reconstructed scattering amplitudes, indicating that
they provide an important step towards unambiguous phase shifts.

We would like to thank the COSY accelerator group for excellent
polarized beam support. This work has been sponsored by the BMBF and
the FZ J\"ulich. 


\begin{thebibliography}{30}
\expandafter\ifx\csname natexlab\endcsname\relax\def\natexlab#1{#1}\fi
\expandafter\ifx\csname bibnamefont\endcsname\relax
  \def\bibnamefont#1{#1}\fi
\expandafter\ifx\csname bibfnamefont\endcsname\relax
  \def\bibfnamefont#1{#1}\fi
\expandafter\ifx\csname citenamefont\endcsname\relax
  \def\citenamefont#1{#1}\fi
\expandafter\ifx\csname url\endcsname\relax
  \def\url#1{\texttt{#1}}\fi
\expandafter\ifx\csname urlprefix\endcsname\relax\def\urlprefix{URL }\fi
\providecommand{\bibinfo}[2]{#2}
\providecommand{\eprint}[2][]{\url{#2}}

\bibitem[{\citenamefont{Machleidt and Slaus}(2001)}]{Machleidt:2001rw}
\bibinfo{author}{\bibfnamefont{R.}~\bibnamefont{Machleidt}} \bibnamefont{and}
  \bibinfo{author}{\bibfnamefont{I.}~\bibnamefont{Slaus}}, \bibinfo{journal}{J.
  Phys.} \textbf{\bibinfo{volume}{G27}}, \bibinfo{pages}{R69}
  (\bibinfo{year}{2001}), \bibinfo{note}{and references herein}.

\bibitem[{\citenamefont{Bedaque and van Kolck}(2002)}]{Bedaque:2002mn}
\bibinfo{author}{\bibfnamefont{P.~F.} \bibnamefont{Bedaque}} \bibnamefont{and}
  \bibinfo{author}{\bibfnamefont{U.}~\bibnamefont{van Kolck}},
  \bibinfo{journal}{Ann. Rev. Nucl. Part. Sci.} \textbf{\bibinfo{volume}{52}},
  \bibinfo{pages}{339} (\bibinfo{year}{2002}), \bibinfo{note}{and references
  herein}.

\bibitem[{\citenamefont{Stoks et~al.}(1993)\citenamefont{Stoks, Klomp,
  Rentmeester, and de~Swart}}]{Stoks:1993tb}
\bibinfo{author}{\bibfnamefont{V.~G.~J.} \bibnamefont{Stoks}},
  \bibinfo{author}{\bibfnamefont{R.~A.~M.} \bibnamefont{Klomp}},
  \bibinfo{author}{\bibfnamefont{M.~C.~M.} \bibnamefont{Rentmeester}},
  \bibnamefont{and} \bibinfo{author}{\bibfnamefont{J.~J.}
  \bibnamefont{de~Swart}}, \bibinfo{journal}{Phys. Rev.}
  \textbf{\bibinfo{volume}{C48}}, \bibinfo{pages}{792} (\bibinfo{year}{1993}).

\bibitem[{\citenamefont{Bystricky et~al.}(1987)\citenamefont{Bystricky,
  {Lechanoine-LeLuc}, and Lehar}}]{Bystricky:1987}
\bibinfo{author}{\bibfnamefont{J.}~\bibnamefont{Bystricky}},
  \bibinfo{author}{\bibfnamefont{C.}~\bibnamefont{{Lechanoine-LeLuc}}},
  \bibnamefont{and} \bibinfo{author}{\bibfnamefont{F.}~\bibnamefont{Lehar}},
  \bibinfo{journal}{J. Phys. (Paris)} \textbf{\bibinfo{volume}{48}},
  \bibinfo{pages}{199} (\bibinfo{year}{1987}).

\bibitem[{\citenamefont{Bystricky et~al.}(1990)\citenamefont{Bystricky,
  {Lechanoine-LeLuc}, and Lehar}}]{Bystricky:1990}
\bibinfo{author}{\bibfnamefont{J.}~\bibnamefont{Bystricky}},
  \bibinfo{author}{\bibfnamefont{C.}~\bibnamefont{{Lechanoine-LeLuc}}},
  \bibnamefont{and} \bibinfo{author}{\bibfnamefont{F.}~\bibnamefont{Lehar}},
  \bibinfo{journal}{J. Phys. (Paris)} \textbf{\bibinfo{volume}{51}},
  \bibinfo{pages}{2747} (\bibinfo{year}{1990}).

\bibitem[{\citenamefont{Nagata et~al.}(1996)\citenamefont{Nagata, Yoshino, and
  Matsuda}}]{Nagata:1996tp}
\bibinfo{author}{\bibfnamefont{J.}~\bibnamefont{Nagata}},
  \bibinfo{author}{\bibfnamefont{H.}~\bibnamefont{Yoshino}}, \bibnamefont{and}
  \bibinfo{author}{\bibfnamefont{M.}~\bibnamefont{Matsuda}},
  \bibinfo{journal}{Prog. Theor. Phys.} \textbf{\bibinfo{volume}{95}},
  \bibinfo{pages}{691} (\bibinfo{year}{1996}).

\bibitem[{\citenamefont{Arndt et~al.}(2000)\citenamefont{Arndt, Strakovsky, and
  Workman}}]{Arndt:2000xc}
\bibinfo{author}{\bibfnamefont{R.~A.} \bibnamefont{Arndt}},
  \bibinfo{author}{\bibfnamefont{I.~I.} \bibnamefont{Strakovsky}},
  \bibnamefont{and} \bibinfo{author}{\bibfnamefont{R.~L.}
  \bibnamefont{Workman}}, \bibinfo{journal}{Phys. Rev. C}
  \textbf{\bibinfo{volume}{62}}, \bibinfo{pages}{34005} (\bibinfo{year}{2000}).

\bibitem[{\citenamefont{Bystricky et~al.}(1998)\citenamefont{Bystricky, Lehar,
  and {Lechanoine-LeLuc}}}]{Bystricky:1998rh}
\bibinfo{author}{\bibfnamefont{J.}~\bibnamefont{Bystricky}},
  \bibinfo{author}{\bibfnamefont{F.}~\bibnamefont{Lehar}}, \bibnamefont{and}
  \bibinfo{author}{\bibfnamefont{C.}~\bibnamefont{{Lechanoine-LeLuc}}},
  \bibinfo{journal}{Eur. Phys. J.} \textbf{\bibinfo{volume}{C4}},
  \bibinfo{pages}{607} (\bibinfo{year}{1998}).

\bibitem[{\citenamefont{Meyer}(1997)}]{meyer97}
\bibinfo{author}{\bibfnamefont{H.~O.} \bibnamefont{Meyer}},
  \bibinfo{journal}{Ann. Rev. of Nucl. and Part. Sci.}
  \textbf{\bibinfo{volume}{47}}, \bibinfo{pages}{235} (\bibinfo{year}{1997}).

\bibitem[{\citenamefont{Haeberli et~al.}(1997)}]{Haeberli:1997}
\bibinfo{author}{\bibfnamefont{W.}~\bibnamefont{Haeberli}}
  \bibnamefont{et~al.}, \bibinfo{journal}{Phys. Rev.}
  \textbf{\bibinfo{volume}{C55}}, \bibinfo{pages}{597} (\bibinfo{year}{1997}).

\bibitem[{\citenamefont{Rathmann et~al.}(1998)}]{Rathmann:1998}
\bibinfo{author}{\bibfnamefont{F.}~\bibnamefont{Rathmann}}
  \bibnamefont{et~al.}, \bibinfo{journal}{Phys. Rev.}
  \textbf{\bibinfo{volume}{C58}}, \bibinfo{pages}{658} (\bibinfo{year}{1998}).

\bibitem[{\citenamefont{von Przewoski et~al.}(1998)}]{vonPrzewoski:1998ye}
\bibinfo{author}{\bibfnamefont{B.}~\bibnamefont{von Przewoski}}
  \bibnamefont{et~al.}, \bibinfo{journal}{Phys. Rev.}
  \textbf{\bibinfo{volume}{C58}}, \bibinfo{pages}{1897} (\bibinfo{year}{1998}).

\bibitem[{\citenamefont{Bourrely et~al.}(1980)\citenamefont{Bourrely, Soffer,
  and Leader}}]{Bourrely:1980mr}
\bibinfo{author}{\bibfnamefont{C.}~\bibnamefont{Bourrely}},
  \bibinfo{author}{\bibfnamefont{J.}~\bibnamefont{Soffer}}, \bibnamefont{and}
  \bibinfo{author}{\bibfnamefont{E.}~\bibnamefont{Leader}},
  \bibinfo{journal}{Phys. Rept.} \textbf{\bibinfo{volume}{59}},
  \bibinfo{pages}{95} (\bibinfo{year}{1980}).

\bibitem[{\citenamefont{Maier}(1987)}]{Maier:1997}
\bibinfo{author}{\bibfnamefont{R.}~\bibnamefont{Maier}},
  \bibinfo{journal}{Nucl.\ Instr.\ and Meth.} \textbf{\bibinfo{volume}{A390}},
  \bibinfo{pages}{1} (\bibinfo{year}{1987}).

\bibitem[{\citenamefont{Eversheim et~al.}(1997)\citenamefont{Eversheim,
  Altmeier, and Felden}}]{eversheim97}
\bibinfo{author}{\bibfnamefont{P.}~\bibnamefont{Eversheim}},
  \bibinfo{author}{\bibfnamefont{M.}~\bibnamefont{Altmeier}}, \bibnamefont{and}
  \bibinfo{author}{\bibfnamefont{O.}~\bibnamefont{Felden}},
  \bibinfo{journal}{Nucl. Phys.} \textbf{\bibinfo{volume}{A626}},
  \bibinfo{pages}{117c} (\bibinfo{year}{1997}).

\bibitem[{\citenamefont{{Albers} et~al.}(1997)}]{albers97}
\bibinfo{author}{\bibfnamefont{D.}~\bibnamefont{{Albers}}} \bibnamefont{et~al.}
  (\bibinfo{collaboration}{EDDA}), \bibinfo{journal}{Phys.\ Rev.\ Lett.}
  \textbf{\bibinfo{volume}{78}}, \bibinfo{pages}{1652} (\bibinfo{year}{1997}).

\bibitem[{\citenamefont{Altmeier et~al.}(2000)}]{Altmeier:2000}
\bibinfo{author}{\bibfnamefont{M.}~\bibnamefont{Altmeier}} \bibnamefont{et~al.}
  (\bibinfo{collaboration}{EDDA}), \bibinfo{journal}{Phys. Rev. Lett.}
  \textbf{\bibinfo{volume}{85}}, \bibinfo{pages}{1819} (\bibinfo{year}{2000}).

\bibitem[{\citenamefont{Bystricky et~al.}(1978)\citenamefont{Bystricky, Lehar,
  and Winternitz}}]{bystricky78}
\bibinfo{author}{\bibfnamefont{J.}~\bibnamefont{Bystricky}},
  \bibinfo{author}{\bibfnamefont{F.}~\bibnamefont{Lehar}}, \bibnamefont{and}
  \bibinfo{author}{\bibfnamefont{P.}~\bibnamefont{Winternitz}},
  \bibinfo{journal}{J. Phys. (Paris)} \textbf{\bibinfo{volume}{39}},
  \bibinfo{pages}{1} (\bibinfo{year}{1978}).

\bibitem[{\citenamefont{Ohlsen}(1973)}]{ohlsen73}
\bibinfo{author}{\bibfnamefont{G.~G.} \bibnamefont{Ohlsen}},
  \bibinfo{journal}{Nucl.\ Instr.\ and Meth.} \textbf{\bibinfo{volume}{109}},
  \bibinfo{pages}{41} (\bibinfo{year}{1973}).

\bibitem[{\citenamefont{Bauer}(2001)}]{Bauer:2001}
\bibinfo{author}{\bibfnamefont{F.}~\bibnamefont{Bauer}}, \bibinfo{type}{Ph.d.
  thesis}, \bibinfo{school}{Universit\"at Hamburg} (\bibinfo{year}{2001}),
  \bibinfo{note}{\url{http://kaa.desy.de/edda/papers/PhDlist.inhalt.html}}.

\bibitem[{\citenamefont{Lehar et~al.}(1987)}]{Lehar:1987pw}
\bibinfo{author}{\bibfnamefont{F.}~\bibnamefont{Lehar}} \bibnamefont{et~al.},
  \bibinfo{journal}{Nucl. Phys.} \textbf{\bibinfo{volume}{B294}},
  \bibinfo{pages}{1013} (\bibinfo{year}{1987}).

\bibitem[{\citenamefont{Allgower et~al.}(2000)}]{Allgower:2000pv}
\bibinfo{author}{\bibfnamefont{C.~E.} \bibnamefont{Allgower}}
  \bibnamefont{et~al.}, \bibinfo{journal}{Phys. Rev.}
  \textbf{\bibinfo{volume}{C62}}, \bibinfo{pages}{064001}
  (\bibinfo{year}{2000}).

\bibitem[{\citenamefont{Allgower et~al.}(2001)}]{Allgower:2001qs}
\bibinfo{author}{\bibfnamefont{C.~E.} \bibnamefont{Allgower}}
  \bibnamefont{et~al.}, \bibinfo{journal}{Phys. Rev.}
  \textbf{\bibinfo{volume}{C64}}, \bibinfo{pages}{034003}
  (\bibinfo{year}{2001}).

\bibitem[{\citenamefont{Perrot et~al.}(1988)}]{Perrot:1988tw}
\bibinfo{author}{\bibfnamefont{F.}~\bibnamefont{Perrot}} \bibnamefont{et~al.},
  \bibinfo{journal}{Nucl. Phys.} \textbf{\bibinfo{volume}{B296}},
  \bibinfo{pages}{527} (\bibinfo{year}{1988}).

\bibitem[{\citenamefont{Fontaine et~al.}(1989)}]{Fontaine:1989ak}
\bibinfo{author}{\bibfnamefont{J.~M.} \bibnamefont{Fontaine}}
  \bibnamefont{et~al.}, \bibinfo{journal}{Nucl. Phys.}
  \textbf{\bibinfo{volume}{B321}}, \bibinfo{pages}{299} (\bibinfo{year}{1989}).

\bibitem[{\citenamefont{Allgower et~al.}(1998)}]{Allgower:1998wf}
\bibinfo{author}{\bibfnamefont{C.~E.} \bibnamefont{Allgower}}
  \bibnamefont{et~al.}, \bibinfo{journal}{Eur. Phys. J.}
  \textbf{\bibinfo{volume}{C1}}, \bibinfo{pages}{131} (\bibinfo{year}{1998}).

\bibitem[{\citenamefont{Jacob and Wick}(1959)}]{JacobWick}
\bibinfo{author}{\bibfnamefont{M.}~\bibnamefont{Jacob}} \bibnamefont{and}
  \bibinfo{author}{\bibfnamefont{G.~C.} \bibnamefont{Wick}},
  \bibinfo{journal}{Annals Phys.} \textbf{\bibinfo{volume}{7}},
  \bibinfo{pages}{404} (\bibinfo{year}{1959}).

\bibitem[{\citenamefont{Conzett}(1994)}]{Conzett:1994rg}
\bibinfo{author}{\bibfnamefont{H.~E.} \bibnamefont{Conzett}},
  \bibinfo{journal}{Rept. Prog. Phys.} \textbf{\bibinfo{volume}{57}},
  \bibinfo{pages}{1} (\bibinfo{year}{1994}).

\bibitem[{\citenamefont{Ball et~al.}(1999{\natexlab{a}})}]{Ball:1999cn}
\bibinfo{author}{\bibfnamefont{J.}~\bibnamefont{Ball}} \bibnamefont{et~al.},
  \bibinfo{journal}{Eur. Phys. J.} \textbf{\bibinfo{volume}{C10}},
  \bibinfo{pages}{409} (\bibinfo{year}{1999}{\natexlab{a}}).

\bibitem[{\citenamefont{Ball et~al.}(1999{\natexlab{b}})}]{Ball:1999yy}
\bibinfo{author}{\bibfnamefont{J.}~\bibnamefont{Ball}} \bibnamefont{et~al.},
  \bibinfo{journal}{Eur. Phys. J.} \textbf{\bibinfo{volume}{C11}},
  \bibinfo{pages}{51} (\bibinfo{year}{1999}{\natexlab{b}}).

\end{thebibliography}

%
\end{document}